\documentclass{revtex-paper}

\newcommand{\onehalf}{\frac{1}{2}}

\DeclareMathSizes{10}{9}{7}{7}

\begin{document}

\title{The $(2+1)$-dimensional Gross-Neveu-Yukawa model at finite temperature, density, and magnetic field within the Functional Renormalization Group}

\author{Justin L.P.~Mauldin}
\email{mauldin@itp.uni-frankfurt.de}
\affiliation{Institute for Theoretical Physics, Goethe University, Max-von-Laue-Str.~1, D-60438 Frankfurt am Main, Germany}

\author{Dirk H.~Rischke}
\email{drischke@itp.uni-frankfurt.de}
\affiliation{Institute for Theoretical Physics, Goethe University, Max-von-Laue-Str.~1, D-60438 Frankfurt am Main, Germany}
\affiliation{Helmholtz Research Academy Hesse for FAIR, Campus Riedberg, Max-von-Laue-Str.~12, D-60438 Frankfurt am Main, Germany}

\begin{abstract}

We investigate the phase diagram of the (2+1)-dimensional Gross-Neveu-Yukawa (GNY) model at finite temperature, density, and magnetic field beyond mean-field, using the Functional Renormalization Group (FRG) in the local potential approximation. 
Large magnetic fields result in magnetic catalysis, a dimensional reduction of the system, and enhancement of chiral symmetry breaking. 
We employ a hydrodynamical algorithm to solve the FRG flow equation for the effective potential, which allows for go further into the infrared region than with previously used methods. 
We find that the chiral condensate exhibits non-trivial behavior in various regions of the phase diagram: several first-order phase transitions and de Haas -- van Alphen oscillations at small magnetic field and large chemical potential, as well as a critical endpoint which shifts to higher temperature with increasing magnetic field.

\end{abstract}

\maketitle

\section{Introduction}\label{Introduction}

Large magnetic fields $qB$ of the order of $10^{14} - 10^{15}$ G (corresponding to $10^{-4} - 10^{-3}\; m_\pi^2$) exist on the surface of neutron stars~\cite{magnetarCoreBFieldGammaRayBursts}, possibly rising to $10^{16} - 10^{19}$ G (i.e., $\sim 10^{-4} - 10\; m_\pi^2$) in their core.
In the early Universe, large magnetic fields of order $10^{18} - 10^{20}$ G (corresponding to $1 - 100\;m_\pi^2$) could have been produced during the electroweak phase transition \cite{Grasso:2000wj} and may persist through the subsequent transition from the Quark-Gluon Plasma to hadronic matter. 
Lastly, short-lived but very large magnetic fields of about $10^{19}$ G (i.e., $\sim 10 \; m_\pi^2$) are present in noncentral heavy-ion collisions \cite{Skokov:2009qp,Tuchin:2013apa}.
How such large magnetic fields influence the properties of strong-interaction matter, and in particular the phase diagram of Quantum Chromodynamics (QCD), is a contemporary topic of great interest~\cite{Andersen:2014xxa, Endrodi:2015oba, Miransky:2015ava, Ferreira:2014kpa, wen_functional_2023, Gao:2026hwr}.

The chiral condensate is an important quantity to distinguish various phases in the QCD phase diagram, in particular it serves as an order parameter for the chiral phase transition~\cite{Pisarski:1983ms}.
At small temperature $T$ and chemical potential $\mu$, the chiral symmetry of QCD is broken, leading to a nonvanishing chiral condensate, while at high $T$ and $\mu$ the chiral symmetry is restored.  
At small $\mu$, the chiral transition is of second order for vanishing quark masses and physical flavor numbers~\cite{Klinger:2026pbe}, while model calculations have found a first-order transition at large $\mu$ and small $T$~\cite{Stephanov:2004wx}.
These two transitions are separated by a tricritical point (TCP), which becomes a second-order critical endpoint (CEP) for nonvanishing quark masses~\cite{Stephanov:1998dy}.
Depending on the concrete values of temperature and chemical potential, a magnetic field can induce either chiral symmetry breaking, a phenomenon known as magnetic catalysis (MC)~\cite{Gusynin_1995}, or restore chiral symmetry, which is known as inverse magnetic catalysis (IMC)~\cite{Preis:2010cq}. 
Furthermore, for given $T$ and $\mu$ it affects the order of the phase transition, for instance by shifting the location of the CEP \cite{Endrodi:2015oba}.

In strong-interaction matter, the IMC effect can occur in two ways, which are distinguished by the underlying mechanism.
In non-abelian gauge theories like QCD, IMC occurs at finite temperature in the form of lowering the chiral transition temperature when the magnetic field is increased~\cite{Bali:2011qj}.
In models for strong-interaction matter, the IMC effect appears at nonzero chemical potential and lowers the chemical potential for the chiral transition \cite{Preis:2010cq, andersen_chiral_2012, Aoki:2015mqa}.

The Gross-Neveu (GN) model for nonvanishing magnetic fields has been studied previously in $(2+1)$ and $(3+1)$ dimensions in mean-field approximation~\cite{Gusynin:1994re}, beyond mean field via Optimized Perturbation Theory~\cite{Kneur:2013cva} and on the lattice~\cite{lenz_magnetized_2023}, as well as within the FRG approach~\cite{PhysRevB.85.195417,Fukushima:2012xw}.
A special feature of the $(2+1)$-dimensional GN model is that a sufficiently large magnetic field induces a condensate even for infinitesimal positive coupling \cite{Gusynin:1994re}. 
In the mean-field approximation, the GN phase diagram exhibits intriguing features: the CEP shifts to higher temperatures with increasing magnetic field, de Haas -- van Alphen oscillations appear in the phase boundary in the $|qB|-\mu$ plane at small values of the magnetic field, and first-order phase transitions can occur already within the chirally broken phase. 
These features were confirmed in Optimized Perturbation Theory at $N_f=2$ \cite{Kneur:2013cva}. 
A (weak) first-order transition was confirmed by lattice studies at $N_f=1$ \cite{lenz_magnetized_2023}.

In this paper, we study the Gross-Neveu-Yukawa (GNY) model, a bosonized version of the GN model \cite{PhysRevD.10.3235}. 
To our knowledge, this model has not yet been studied at nonvanishing magnetic field, neither in $(2+1)$ nor in $(3+1)$ dimensions.
Our present study aims to fill this gap for $(2+1)$ dimensions.
The extension to $(3+1)$ dimensions will be left to future work.

While relatively simple, the GN and GNY models share two key features with QCD: spontaneous chiral symmetry breaking and asymptotic freedom. 
Additionally, while the GN model is renormalizable in $d<3$ spatial dimensions, the GNY model is also renormalizable in $d=3$ dimensions.
This allows to study magnetic-field effects independently of the ultraviolet (UV) cutoff.

In order to investigate the phase diagram of the GNY model, we employ the functional renormalization group (FRG)~\cite{Wetterich:1992yh} in local potential approximation (LPA) and compute the corresponding effective potential as a function of $T$ and $\mu$.
The restriction to two spatial dimensions allows us to choose larger UV cutoffs and go further into the infrared (IR) region as for three spatial dimensions.
We use a hydrodynamic finite-volume approach~\cite{Grossi:2019urj,Stoll2021arXiv210810616S} to solve the FRG flow equation for the derivative of the effective potential, which greatly improves the accuracy of the numerical solution as compared to previously employed finite-difference methods.
Our main results are that we qualitatively confirm the above mentioned features of the phase diagram found in Refs.~\cite{Kneur:2013cva,lenz_magnetized_2023}, i.e., de Haas -- van Alphen oscillations of the phase boundary in the $|qB| - \mu$ plane and several first-order transitions within the chirally broken phase at small chemical potentials, as well as a shift of the CEP to higher temperatures with increasing magnetic field.
However, with our hydrodynamic approach to solve the FRG flow equation for the effective potential, we resolve them quantitatively with much higher precision.

This paper is organized as follows.
In Sec.~\ref{sec:FRG} we present the FRG flow equations for the effective potential at finite temperature and chemical potential, for both zero and nonzero magnetic field. 
In Sec.~\ref{sec:results} we discuss our results for the phase diagram as function of temperature, chemical potential, and magnetic field. 
Conclusions and an outlook are given in Sec.~\ref{sec:conclusion}.
A detailed derivation of the FRG flow equations for zero and nonzero magnetic field is relegated to the Appendix.
There we also describe the numerical scheme used to solve the flow equations. 

\section{The Gross-Neveu-Yukawa model with the Functional Renormalization Group}
\label{sec:FRG}

The GNY model~\cite{Zinn-Justin:1991ksq} is derived from the GN model~\cite{PhysRevD.10.3235} in two steps.
First, the four-fermion interaction is bosonized using a Hubbard-Stratonovich transformation, which couples the fermions to a scalar field $\varphi$ via a Yukawa interaction of coupling strength $h$. 
Finally, the scalar field is made dynamical by inserting a kinetic term for the scalar field by hand, which allows to include bosonic fluctuations. 
The Lagrangian in Euclidean space-time is then given by 
\begin{equation}
    \mathcal{L}
    =
    \Bar{\psi}
    \left(
    \slashed{\partial}
    +
    \frac{h}{\sqrt{N_f}}\varphi
    \right)
    \psi
    -
    \frac{1}{2}
    \varphi\, \Box \varphi  
    +
    \frac{h^2}{2 g^2}\,\varphi^2
    \;,
\end{equation}
where $N_f$ is the number of fermion flavors and $g^2$ the four-fermion coupling. 
The factor $1/\sqrt{N_f}$ in the Yukawa coupling arises because the four-fermion interaction in the GN model is proportional to $g^2/(2N_f)$.
At nonzero temperature, the $(2+1)$-dimensional momentum-space integral is replaced by a 2-dimensional integration over spatial momentum and a Matsubara sum,
\begin{align}
    \int \frac{\mathrm{d}^3 p}{(2\pi)^3} &\rightarrow 
    \frac{1}{\beta}
    \sum_{n=-\infty}^{\infty}
    \int \frac{\mathrm{d}^2 p}{(2\pi)^2} 
    \;.
\end{align}
At nonzero temperature and chemical potential, the Euclidean energy is replaced by $p_0 \rightarrow \nu_n +  i \mu$ for fermions, and $p_0 \rightarrow \omega_n$ for bosons. 
The fermionic and bosonic Matsubara frequencies are given by $\nu_n = (2n + 1) \pi T$ and $\omega_n = 2 n \pi T$, respectively.

The FRG approach is a powerful method to determine  the full quantum effective action $\Gamma[\Phi]$.
Starting with the classical action $\mathcal{S}[\Phi]$ as initial condition at a sufficiently large UV momentum scale $\Lambda$, solving a flow equation in the RG momentum scale $k$ one successively integrates out momentum shells to arrive at $\Gamma[\Phi]$ in the IR. 
Here, we are using the Wetterich form of the FRG flow equation~\cite{Wetterich:1992yh}:
\begin{equation}
    \label{eq:wetterich}
    \partial_t {\Gamma}_t [\Phi] = 
    \frac{1}{2} \mathrm{STr}\left[ 
        \frac{
            \left(    \partial_t R_t \right)
        }{
            {\Gamma}_t^{(2)}[\Phi] + R_t
        }
    \right]
    \;,
\end{equation}
where $\Phi$ is a superfield containing all the fields of our model and $R_t$ is a regulator function. 
The supertrace is a trace resp.~integral over all internal fields, indices, and momenta. 
In addition, the trace over fermionic fields gets a minus sign. 
Furthermore, we replaced the RG scale $k$ with the RG time $t \equiv - \ln (k/\Lambda)$. 
The bosonic and fermionic regulators are given by
\begin{align}
    \label{eq:regulatorBoson}
    R_b (t, p) &= {\vec{p}}^{\;2} r_b (t, p) \;,\\
    \label{eq:regulatorFermion}
    R_f (t, p) &= - i \vec{p}\cdot \vec{\gamma} \; r_f(t, p) \;,
\end{align}
where we parametrized the regulators in terms of the bosonic and fermionic regulator shape functions $r_b(t, p)$ and $r_f(t, p)$. 
The regulators act as scale-dependent mass terms, suppressing momenta larger than the RG scale $k$; the derivative of the regulator implements Wilson's idea of integrating out momentum shells.

In general, Eq.~(\ref{eq:wetterich}) cannot be solved in full generality, because the right-hand side involves the two-point vertex function $\Gamma_t^{(2)}[\Phi]$, which itself obeys a flow equation involving higher-point functions.
These again have to be determined by their respective flow equations involving ever higher $n$-point functions.
In practice, in order to solve the Wetterich equation this infinite tower has to be truncated.
To this end, we employ the derivative expansion in the so-called local potential approximation (LPA), i.e., the effective average action at the scale $k$ is given by
\begin{align}
    \label{gny:LPA_coordinate}
    {\Gamma}_t \left[
        \Bar{\psi}, \psi, \varphi
    \right]
    =
    &
    \int \mathrm{d}^2 x 
    \int_{0}^{\beta} \mathrm{d}\tau
    \left[ 
        \Bar{\psi}\left( \slashed{\partial} - \mu \gamma^4 + \frac{h}{\sqrt{N_f}} \varphi\right)\psi
    \right.\nonumber\\
    &\left.
        -
    \frac{1}{2} \varphi \,\Box \varphi + U(t, \varphi)
    \right]
    \;,
\end{align}
where the effective potential $U(t, \varphi)$ is the only quantity which depends on the RG scale $k= \Lambda\, e^{-t}$. 

For the regulator shape functions, we choose flat Litim regulators for spatial momenta~\cite{Litim:2001up}
\begin{align} \label{eq:shape_b}
    r_b (t, p)
    &=
    \left(\frac{k^2}{\vec{p}^{\,2}} - 1\right) 
    \theta \left(1 - \frac{p^2}{k^2}\right)
    \;,
    \\
    r_f (t, p)
    &=
    \left(\frac{k}{p} - 1\right)
    \theta\left(1 - \frac{p^2}{k^2}\right)
    \;. \label{eq:shape_f}
\end{align}

Furthermore, we evaluate both flow equations on a constant background
\begin{equation}
    \label{eq:constantBackground}
    \varphi(\tau, \vec{x}) = \sigma
    \;, \quad
    \bar{\psi}(\tau, \vec{x}) = \psi(\tau, \vec{x}) = 0\;.
\end{equation}

Inserting the truncation  (\ref{gny:LPA_coordinate}) into the Wetterich equation \eqref{eq:wetterich} and evaluating on the background \eqref{eq:constantBackground} yields a flow equation for the effective potential:
\begin{equation}
    \partial_t U(t, \sigma)
    =
    \begin{tikzpicture}[scale=0.5, baseline={(current bounding box.center)}]
      \draw[decorate, decoration={snake, segment length=4pt, amplitude=2pt}] (0,0) circle (1cm);
      \draw[cross] (0,1) circle (0.3cm);
     \node[below] at (0,-1.2) {};
    \end{tikzpicture}
    -
    2
    \,\,
    \begin{tikzpicture}[scale=0.5, baseline={(current bounding box.center)}]
      \draw (0,0) circle (1cm);
      \draw[->] (0,-1) -- (-0.01,-1);
      \draw[fill=white, cross] (0,1) circle (0.3cm);
     \node[below] at (0, -1.2) {};
    \end{tikzpicture}
    \;.
\end{equation}

The right-hand side of the flow equation consists of a loop for the scalar field (denoted by a wavy line) and the other for the fermionic fields (denoted by a full line).
The derivation for the flow equations is given in Appendix \ref{Flow-Equations}. 
In the next two subsections we will analyze the flow equation separately for zero and for nonzero external magnetic field.

\subsection{Zero Magnetic Field}

As shown in Appendix \ref{Flow-Equations}, in $2+1$ dimensions the effective potential for vanishing external magnetic field reads
\begin{widetext}
\begin{equation}
\label{eq:twoDimWithoutExtB}
    \partial_t U(t, \sigma)
    =
    \frac{d_\gamma }{8 \pi}
    \frac{k^{4}}{ E_f (t, \sigma)}
    \left[
        1
        -
        n_f \left(\beta\left[E_f (t, \sigma)+ \mu\right] \right)
        -
        n_f \left(\beta\left[E_f (t, \sigma)- \mu\right]\right)
    \right]
    -
    \frac{d_\gamma }{8 \pi}
    \frac{k^{4}}{N_f E_b (t, \sigma)}
    \left[
        1 + 2 n_b \left(\beta E_b(t, \sigma)\right)
    \right]
    \;,
\end{equation}
\end{widetext}
where $n_f(x)$ and $n_b(x)$ are the Fermi-Dirac and Bose-Einstein distribution functions, respectively. 
The dimension of the matrix representation of the Clifford algebra is denoted by $d_\gamma$ (for the value of $d_\gamma$, see discussion below).
The fermionic and bosonic energies are given by 
\begin{align}
    E_f (t, \sigma)
    & =
    \sqrt{k^2 + (h \sigma)^2}
    \;, \\    E_b (t, \sigma)
    & =
    \sqrt{k^2 + \partial_\sigma^2 U(t, \sigma)}
    \;.
\end{align}
One observes that the Yukawa coupling generates an effective mass for the fermions $m_f = h\sigma$, while the squared curvature mass of the scalar field is given by $m_\sigma^2 = \partial_\sigma^2 U(t, \sigma)$.

\subsection{Nonzero Magnetic Field}

In $(2+1)$ dimensions and zero magnetic field, there are two equivalent irreducible representations of the Clifford algebra. 
For nonzero magnetic field, these representations are no longer equivalent and favor either particles or antiparticles \cite{shovkovy_magnetic_2013}.
To avoid this we work in a reducible representation of the Clifford algebra; in our case the one where the dimension of the Dirac matrices is $d_\gamma = 4$. 
We choose the external magnetic field to point along the $z$-axis, such that the vector potential can be chosen as $A_\mu = (0, 0, B x, 0)$.

In the following we will consider the particles to move exclusively in the 2-dimensional $(xy)$-plane, i.e., their momentum component in direction of the magnetic field vanishes $p_z = 0$. 
Solving the Dirac equation in a constant external magnetic field $qB$ yields the following eigenvalues for fermions\footnote{For the sake of simplicity we assume all fermion flavors to carry the same electric charge $q$. Otherwise, the $SU(N_f)$ flavor symmetry is broken, which necessitates the introduction of separate chiral condensates for the different fermion flavors~\cite{Gao:2026hwr}.},
\begin{equation}
\label{eq:DispersionRelationLandauLevels}
    \lambda_{l, s_z}^2
    =
    m_f^2 + \nu_n^2 + (2l + 1 - 2s_z) |q B|
    \;,
\end{equation}
where $l\geq 0$ is the orbital angular-momentum quantum number and $s_z = \pm 1/2$ is the spin projection along the $z$-axis. 
As one observes, due to the fact that the fermionic motion is restricted to the plane transverse to the magnetic field, the eigenvalues do not depend on $p_z$.
Note that the eigenvalue for $l+1$ and $s_z=1/2$ becomes degenerate with the one for $l$ and $s_z=-1/2$.
Only the level $(l,s_z) =(0,1/2)$ has degeneracy one.
In the following, we label the Landau levels by the variable $l'\equiv l-s_z + 1/2$, where the lowest Landau level $l'=0$ has degeneracy one and all Landau levels $l'>0$ have degeneracy two. 
Consequently, for a nonzero magnetic field, one can simply perform the following replacements in the previous calculation for vanishing magnetic field,
\begin{align}
    \label{eq:momentumReplacement}
    \vec{p}^{\,2} & \rightarrow 2 |q B| l \;,\\
\label{eq:volumeIntegralToLandauSum}
    \int\frac{\mathrm{d}^2 p }{(2\pi)^2}
   & \rightarrow 
    \frac{|qB|  }{4\pi}
        \sum_{l=0}^{\infty}
        \alpha_l
        \;,
\end{align}
where we have replaced $l' \rightarrow l$ for notational simplicity and $\alpha_l = 2 - \delta_{0l}$ is the degeneracy of the respective Landau level. 
This dimensional reduction leads to MC~\cite{Gusynin:1994re}.

Since the scalar field carries no electric charge, the boson loop remains unaffected by a nonzero magnetic field.
Using the Litim regulator shape function the sum over Landau levels is truncated and can be trivially performed (for details, see Appendix~\ref{Flow-Equations}),
\begin{align}
    \label{eq:trivialSum}
    \sum_{l=0}^{\infty}
    \theta \left( k^2 - 2 |q B| l \right)
    =
    \sum_{l=0}^{N_{LL}(k^2)}
    = 
    N_{LL}(k^2)
    \;,
\end{align}

\noindent where $N_{LL} (k^2) =\left \lfloor \frac{k^2}{2 | q B|}\right \rfloor$ is the number of Landau levels contributing to the flow at the scale $k$. 
The flow equation for the effective potential in $2 + 1$ dimensions in an external magnetic field finally reads

\begin{widetext}
\begin{equation}
\label{eq:TwoPlusOneMagneticFlowEquation}
    \partial_t U(t, \sigma)
    =
        \frac{
        d_\gamma  |q B| k^2 \left[1 + 2N_{LL}(k^2)\right]
    }{
        8    \pi E_f (t, \sigma)
    }
    \left[
        1
        -
        n_f \left(\beta \left[E_f(t, \sigma) + \mu \right]\right)
        -
        n_f \left(\beta \left[E_f(t, \sigma) - \mu \right]\right)
    \right]
    - \frac{k^4}{8 \pi N_f E_b(t, \sigma)}
    \left[
        1
        +
        2 n_b (\beta E_b(t, \sigma))
    \right]
    \;.
\end{equation}
\end{widetext}
We note that for $k < \sqrt{2 |q B|}$, only the lowest Landau level (LLL) contributes, rendering the LLL approximation exact in the IR (for nonzero $qB$) and dominating the contribution to the flow\footnote{This can be taken as a justification for the frequently used LLL approximation. If one uses a regulator shape function with a smooth cutoff, the transition will be more gradual.}.
For $k \gg |qB|$, the floor function can be approximated by its argument, reducing Eq.~\eqref{eq:TwoPlusOneMagneticFlowEquation} to Eq.~\eqref{eq:twoDimWithoutExtB}. 
In the UV, the contribution of the magnetic field is therefore negligibly small.

    \section{Results}
\label{sec:results}    
    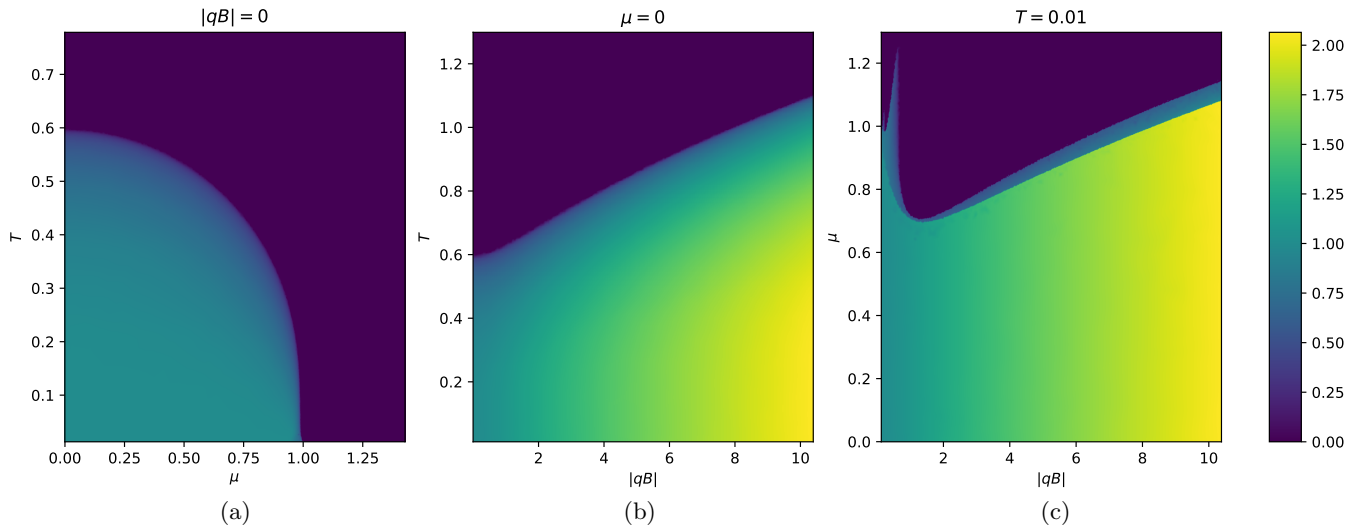
\begin{figure*}[t!]
    \centering
    \begin{tikzpicture}
        \node (img) {\includegraphics[width=\textwidth]{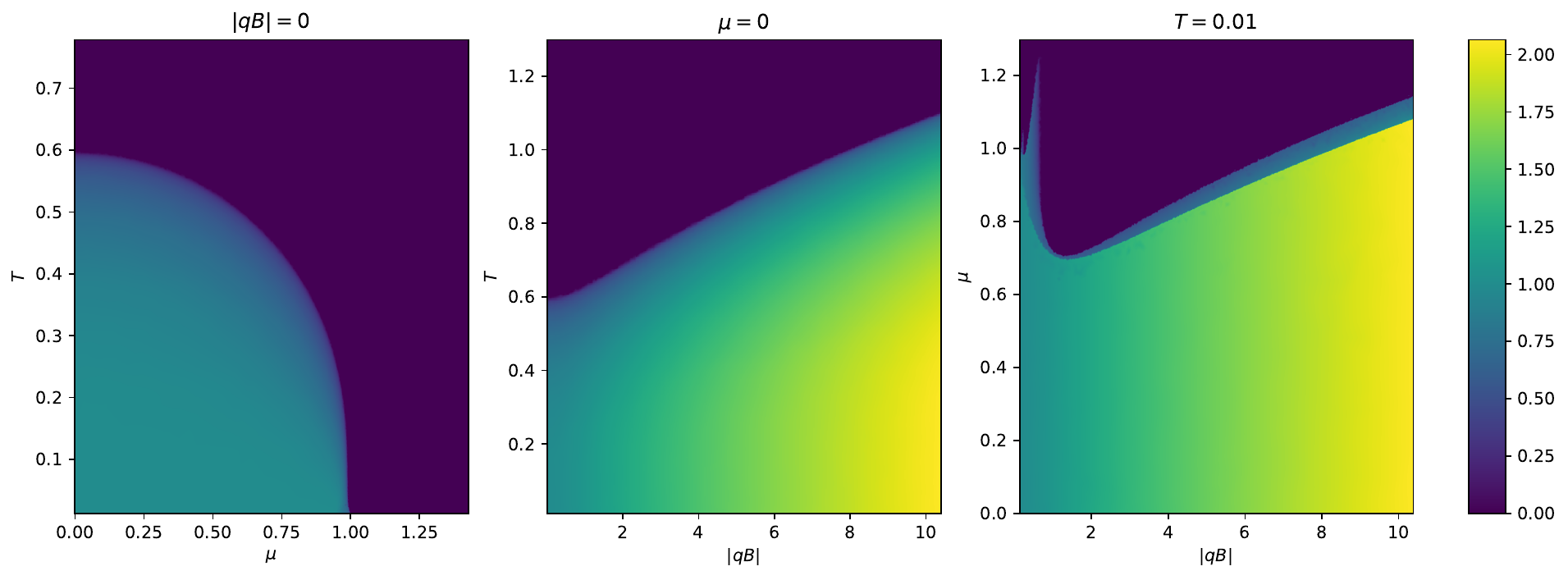}};
        
        \node at (-5.85,-3.5) {(a)};
        \node at ( -0.5,-3.5) {(b)};
        \node at ( 5.,-3.5) {(c)};
    
    \end{tikzpicture}
    \caption{\label{fig:threePdsProjections} Contour plots of the chiral condensate in (a) the $T$--$\mu$ plane at vanishing magnetic field, (b) the $T$--\(|qB|\) plane at $\mu=0$, and (c) the $\mu$--\(|qB|\) plane at $T=0.01$. All results are for $N_f=2$ flavors and a UV cutoff of $\Lambda=10^3$.}
\end{figure*}

\begin{figure*}[t!]
    \begin{subfigure}{0.32\textwidth}
        \includegraphics[width=1.1\linewidth]{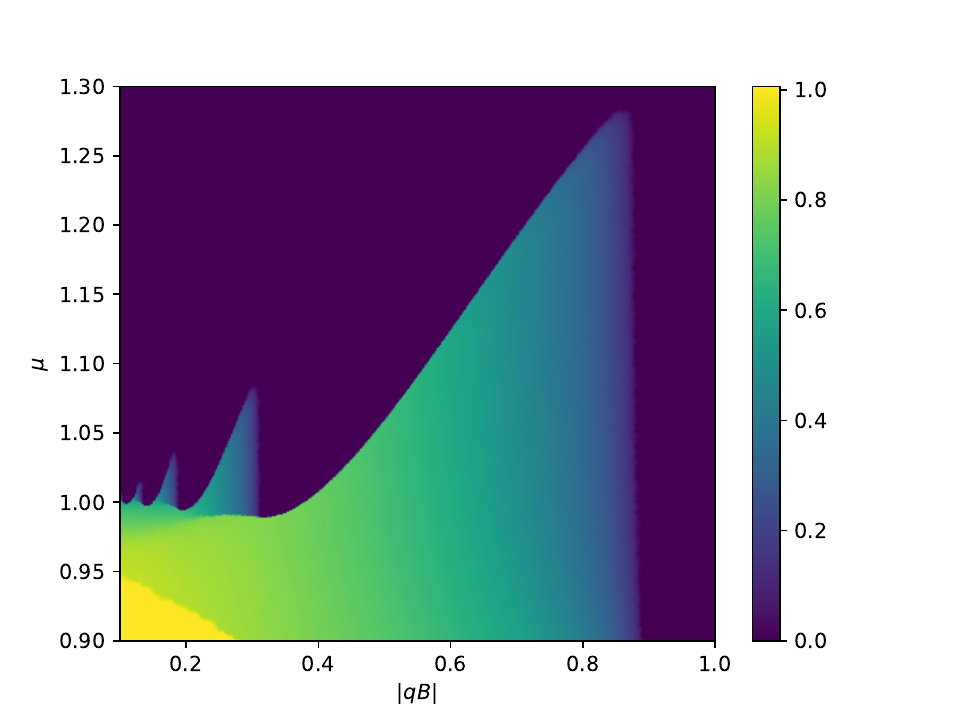}
        \caption{}
        \label{fig:a}
    \end{subfigure}
    \begin{subfigure}{0.32\textwidth}
        \includegraphics[width=1.1\linewidth]{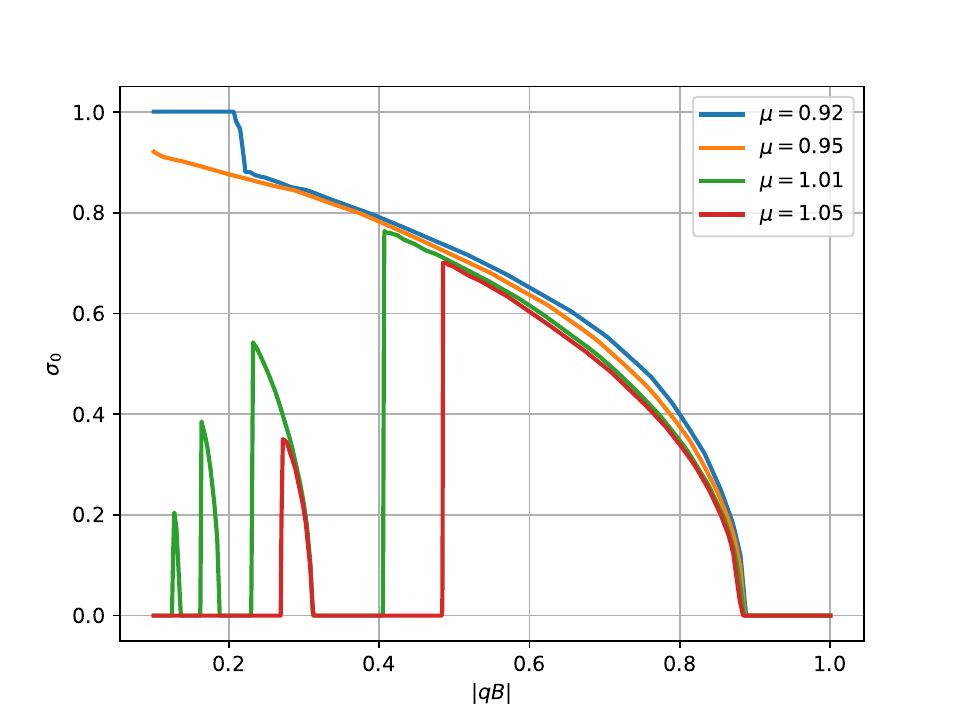}
        \caption{}
        \label{fig:b}
    \end{subfigure}
    \begin{subfigure}{0.32\textwidth}
        \includegraphics[width=1.1\linewidth]{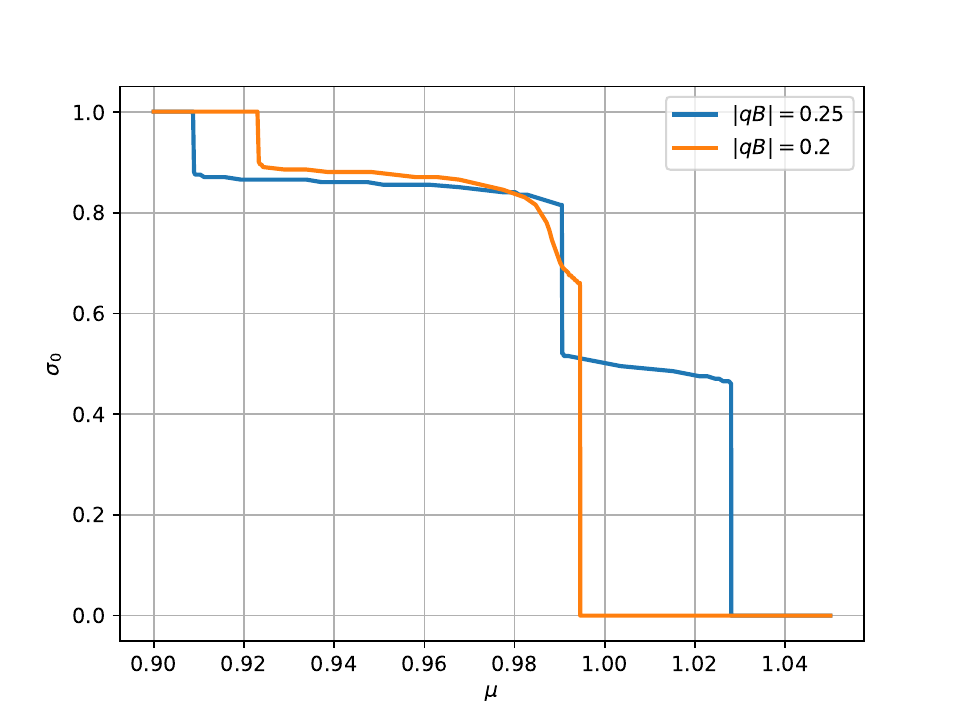}
        \caption{}
        \label{fig:c}
    \end{subfigure}
  \caption{\label{fig:pdMuSmallB} (a) Close-up of the phase diagram in the $\mu$--\(|qB|\) plane at small \(|qB|\) and large $\mu$ ($T=0.01$, $N_f=2$, $\Lambda=10^3$). 
  (b) Chiral condensate as a function of $|qB|$ at $\mu=0.92$ (blue line), $\mu=0.95$ (orange line), $\mu = 1.01$ (green line), and $\mu =1.05$ (red line).
   (c) As in (b), as a function of $\mu$ at  \(|qB|=0.2\) (orange line) and $|qB|=0.25$ (blue line).}
\end{figure*}

We solve the flow equations by using the Kurganov-Tadmor scheme~\cite{KURGANOV2000241}, which allows us to compute the effective potential in high resolution and thus capture non-trivial features of the phase diagram (see Introduction). 
We relegate a more detailed discussion of the numerical solution scheme to Appendix~\ref{sec:numerics}. 
All dimensionful quantities are given in units of $h \sigma_0$, where $\sigma_0$ is the vacuum expectation value of the chiral condensate and we set $h=1$ for the sake of convenience. 
For high-resolution representations of the various projections of the phase diagram, we used \textit{adaptive} \cite{Nijholt2019}. 

In Fig.~\ref{fig:threePdsProjections} we show the projections of the phase diagram onto the (a) $T$--$\mu$, (b) $T$--\(|qB|\), and (c) $\mu$--\(|qB|\) planes for $N_f=2$ with a cutoff of $\Lambda=10^3$. 
At zero magnetic field, Fig.~\ref{fig:threePdsProjections} (a), the chiral transition is almost entirely of second order, with a small first-order region at low $T$ and a TCP that moves towards $T=0$ with increasing UV cutoff.

In the $T$--\(|qB|\) plane at zero chemical potential, Fig.~\ref{fig:threePdsProjections} (b), we find MC with a monotonically increasing critical temperature.
As in other model calculations~\cite{Bandyopadhyay_2021,FRAGA2014154}, the phase transition remains of second order, and in order to obtain IMC at nonzero $T$, one has to introduce a magnetic field-dependent coupling.

More interesting features appear in the $\mu$--$|qB|$ plane of the phase diagram, Fig.~\ref{fig:threePdsProjections} (c).
At the small temperature $T=0.01$,  MC is dominant at small chemical potentials, i.e., the condensate grows monotonically with \(|qB|\).
On the other hand, at larger $\mu$, we first observe IMC at small values of $|qB|$, i.e., the chemical potential at the second-order phase transition to the restored phase becomes smaller.
Above $|qB| \simeq 1$, MC sets in and the chemical potential at the transition increases with the magnetic field.

The chiral transition in the small-$|qB|$ region of Fig.~\ref{fig:threePdsProjections} (c) is accompanied by the appearance of de Haas -- van Alphen oscillations.
Figure~\ref{fig:pdMuSmallB} (a) shows a close-up of this particular region of the $\mu$--$|qB|$ plane.
In Fig.~\ref{fig:pdMuSmallB} (b) we show the chiral condensate as a function of $|qB|$ for several values of the chemical potential.
For $\mu=0.92$, a first-order transition can be observed at $|qB| \simeq 0.2$.
For $\mu = 0.95$, the transition is of second order.
On the other hand, at $\mu=1.01$ and $\mu=1.05$, when increasing $|qB|$ the de Haas -- van Alphen oscillations manifest themselves in a sequence of first-order transitions into the broken phase, followed by second-order symmetry-restoring second-order transitions.
The first-order transitions can also be observed when plotting the chiral condensate as a function of $\mu$ for fixed $|qB|$, cf.~Fig.~\ref{fig:pdMuSmallB} (c).

\begin{figure*}[t!]
    \centering
    \begin{tikzpicture}
        \node (img) {\includegraphics[width=\textwidth]{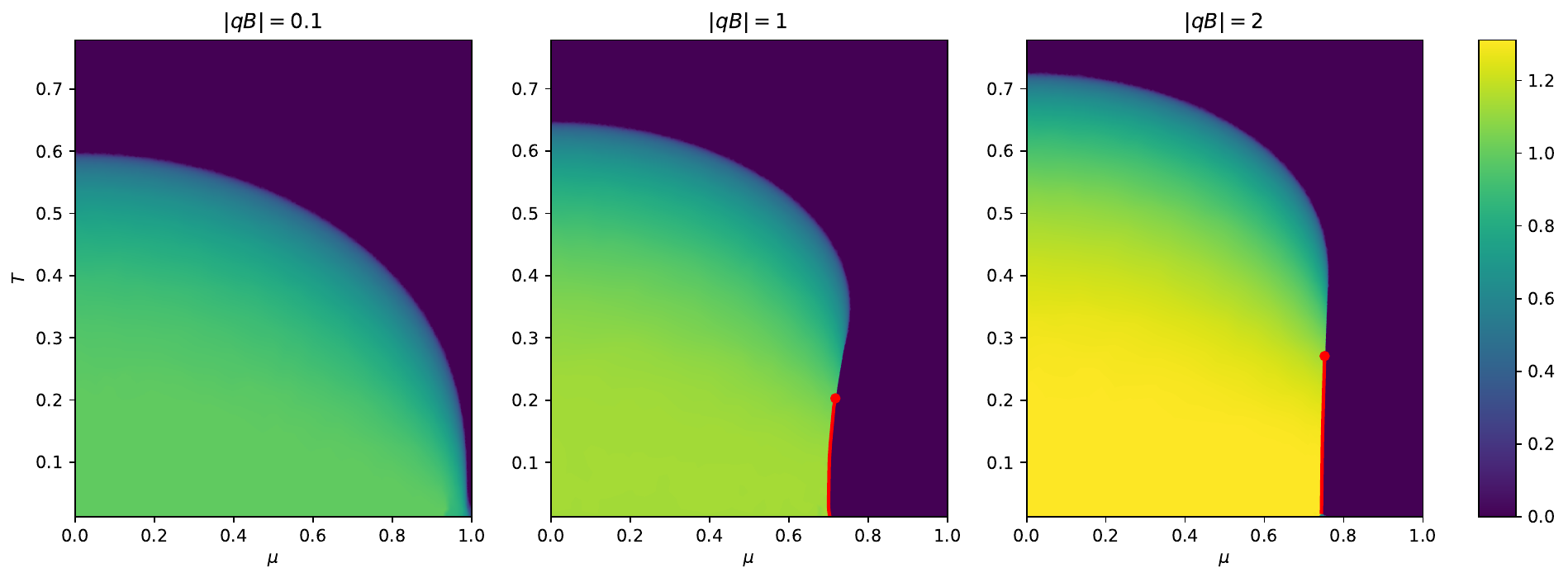}};
        
        \node at (-5.85,-3.5) {(a)};
        \node at ( -0.5,-3.5) {(b)};
        \node at ( 5.,-3.5) {(c)};
    
    \end{tikzpicture}
    \caption{\label{fig:threePdsIncreasingB} Contour plots of the chiral condensate in the $T$--$\mu$ plane at (a) $|qB| = 0.1$, (b) $|qB|=1$, and (c) $|qB| =2$ ($N_f=2$, $\Lambda=10^3$). First-order transition line
    and TCP are marked in red.}
\end{figure*}

In Fig.~\ref{fig:threePdsIncreasingB} we show the phase boundaries in the $T$--$\mu$ plane at three different magnetic-field strengths.
Comparing Figs.~\ref{fig:threePdsIncreasingB} (a) -- (c), we first confirm MC at vanishing chemical potential, i.e., the critical temperature for chiral symmetry restoration grows with increasing magnetic field, cf.~Fig.~\ref{fig:threePdsProjections} (b).
At low \(|qB|\), Fig.~\ref{fig:threePdsIncreasingB} (a), the phase diagram is almost identical to the zero-field case. 
Turning on \(|qB|\), Figs.~\ref{fig:threePdsIncreasingB} (b) and (c), has two notable effects. 
First, a first-order chiral transition line emerges at small temperatures, which elongates as we increase the magnetic field, shifting the TCP to higher temperatures. 
Second, it produces a back-bending of the phase boundary due to IMC, lowering the chemical potential at the transition point. 
This is distinct from the back-bending of the transition line in the quark–meson model at zero magnetic field~\cite{Schaefer:2004en,Tripolt:2017zgc}.
The back-bending observed here is already present in the mean-field approximation.
As discussed above (Fig.~\ref{fig:threePdsProjections} (c)), MC ultimately dominates at even larger \(|qB|\), and the critical chemical potential grows again with \(|qB|\), while the back-bending disappears.

\section{Conclusions}
\label{sec:conclusion}
In this work, we investigated the phase diagram of the (2+1)-dimensional Gross-Neveu-Yukawa model at finite temperature, chemical potential, and magnetic field beyond the mean-field approximation.
We employed the FRG approach in LPA and solved the FRG flow equation via a hydrodynamic scheme to obtain the effective potential.

We found that, at zero chemical potential, MC dominates and the chiral condensate increases monotonically with \(|qB|\). 
Nevertheless, in the low-$T$, intermediate-$\mu$ region of the phase diagram we also observed IMC for intermediate values of the magnetic field. 
The phase diagram exhibits a rich structure: de Haas -- van Alphen oscillations at small \(|qB|\) and large $\mu$, a TCP that shifts to higher $T$ and lower $\mu$ with increasing \(|qB|\), 
and a characteristic back-bending of the phase boundary, which is distinct from the back-bending effect in the quark–meson model at zero magnetic field. 
These features were previously also observed in the Gross-Neveu model in the mean-field approximation~\cite{lenz_magnetized_2023}
and in Optimized Perturbation Theory~\cite{Kneur:2013cva}.
Here, we show that they also exist in the Gross-Neveu-Yukawa model. 
The hydrodynamic scheme that we employed to solve the FRG flow equation allows to resolve these features with unprecedented precision.

As an outlook, one could extend the present investigation to (3+1) space-time dimensions and include pions, to arrive at the quark-meson model. 
We anticipate that we can improve earlier calculations in the framework of that model~\cite{Andersen:2012bq,Skokov:2011ib} by a better resolution of the deep-IR region in the FRG flow. 

\section*{Acknowledgments}
The authors thank Chowdhury Aminul Islam, Ashutosh Dash, Keiwan Jamaly, Lutz Kiefer, and Johannes P\"oplau for valuable discussions. J.L.P.M.~acknowledges support by the GSI F\&E program.
This work is supported in part
by the Deutsche Forschungsgemeinschaft (DFG, German Research Foundation) through
the CRC-TR 211 “Strong-interaction matter under extreme conditions” – project number
315477589–TRR 211.
\appendix
\section{Flow Equations}
\label{Flow-Equations}

In this appendix, we explicitly derive the flow equations in the case of zero and nonzero magnetic field.

\subsection{Zero Magnetic Field}
In LPA, only the effective potential $U$ depends on the RG scale, such that the Wetterich equation  \eqref{eq:wetterich} reads
\begin{align}
    \lefteqn{\mathcal{V}_2
    \partial_t U(t, \sigma)
    =
        \left. \partial_t {\Gamma}_t[\Bar{\psi}, \psi, \varphi] \right|_{\varphi = \sigma, \Bar{\psi}=\psi = 0}}
    \nonumber\\
    &=  \left.
    \text{STr} \left[ 
        \left(\frac{1}{2} \partial_t R_t \right) 
        \left({\Gamma}_{t}^{(2)}[\Phi] + R_t \right)^{-1}
    \right]
    \right|_{\varphi = \sigma, \Bar{\psi}= 0, \psi = 0}
    \;,
\end{align}
where $\mathcal{V}_2 = \beta (2\pi)^2 \delta^{(2)}(0)$ is the two-dimensional volume element.

For vanishing magnetic field, 
Fourier-transforming the right-hand side of Eq.~(\ref{gny:LPA_coordinate}) gives
\begin{align}
\label{gny:LPA_momentum}
    &\Gamma_t [\bar{\psi}, \psi, \varphi]
    \nonumber \\
    &
    =
    \frac{1}{\beta}
    \sum_{n=-\infty}^{\infty}\int \frac{\mathrm{d}^2 p}{(2\pi)^2}
    \left\{
        \bar{\psi}
        \left[
            i\Vec{p} \cdot \Vec{\gamma}
            +
            i\left(\nu_n + i\mu\right) \gamma^4 
            + 
            \frac{h}{\sqrt{N_f}}\varphi
        \right]
        \psi
    \right.
    \nonumber \\
    &\left.
        +
        \frac{1}{2}
        \varphi
        \left(
            \omega_n^2
            +
            \Vec{p}^{\,2}
        \right)
        \varphi
        +
        U(t, \varphi)
    \right\}
    \;.
\end{align}
In this particular truncation, the two-point vertex function is
\begin{align}
    {\Gamma}_t^{(2)}
    =
    &
    \begin{pmatrix}
        {\Gamma}_t^{\varphi \varphi}  & 0 & 0 \\
        0   & 0     & {\Gamma}_t^{\Bar{\psi} \psi}\\
        0   & {\Gamma}_t^{\psi \Bar{\psi}} & 0
    \end{pmatrix}
    \;,
\end{align}
where
\begin{align}
    {\Gamma}^{\varphi \varphi}_t
    &=
    \omega_n^2 + \Vec{p}^{\,2}
    +
    \partial_\sigma^2 U(t, \sigma)
    \;,
    \\
    {\Gamma}^{\Bar{\psi} \psi}_t
    &=
    -
    i
    \left[
        \left(\nu_n + \mu\right)\gamma^{4}
        +
        \Vec{p}\cdot \Vec{\gamma}
        +
        \frac{h}{\sqrt{N}}\varphi
    \right] \equiv -\Gamma^{\psi\Bar{\psi}}_t
    \;.
\end{align}
The regulator matrix is given by
\begin{align}
    R_t
    =
    &
    \begin{pmatrix}
        R_{t \, b} & 0 & 0 \\
        0 & 0 & R_{t \, f} \\
        0 & -R_{t \, f} & 0
    \end{pmatrix}\;.
\end{align}

Adding the two-point vertex and the  respective regulator functions, we obtain
\begin{align}
    & {\Gamma}^{\varphi \varphi}_t
    +
    R_b
    =
    \omega_n^2 + \Vec{p}^{\,2}
    \left[
        1 + r_b(t, p)
    \right]
    +
    \partial_\sigma^2 U(t, \sigma)
    \;,
    \\
 & {\Gamma}^{\Bar{\psi} \psi}_t
    +
    R_f = \nonumber \\
   &  =
    -
    \left\{
        i
        \left(\nu_n + \mu\right)\gamma^4
        +
        i
        \Vec{p}\cdot \Vec{\gamma}
        \left[
            1
            +
            r_f (t, p)
        \right]
        +
        \frac{h}{\sqrt{N_f}}\varphi
    \right\}
    \;.
\end{align}
Inverting these yields
\begin{align}
\label{eq:inverseTwoPointFunctionBosons}
    \left(
        {\Gamma}^{\varphi \varphi}_t
        +
        R_b
    \right)^{-1}
    & =
       \frac{1}{
        \omega_n^2 + \Vec{p}^{\,2}
        \left[
            1 + r_b(t, p)
        \right]
        +
        \partial_\sigma^2 U(t, \sigma)
    }
    \;,
    \\
\label{eq:inverseTwoPointFunctionFermions}
    \left(
        {\Gamma}^{\Bar{\psi} \psi}_t
        +
        R_f
    \right)^{-1}
    & =
    \frac{
        i\left(\nu_n + \mu\right)\gamma^4
        +
        i\Vec{p}\cdot \Vec{\gamma}
        \left[
            1
            +
            r_f (t, p)
        \right]
        -
        m_f
    }{
        (\nu_n + i \mu)^2 + \Vec{p}^{\,2} \left[1 + r_f (t, p)\right]^2
        +
        m_f^2
    }
    \;.
\end{align}

Equations~(\ref{eq:inverseTwoPointFunctionBosons}), (\ref{eq:inverseTwoPointFunctionFermions}) are then inserted together with their respective regulator shape functions (\ref{eq:shape_b}), (\ref{eq:shape_f}) into the Wetterich equation \eqref{eq:wetterich} evaluated on the background-field configuration~\eqref{eq:constantBackground}. 
The boson loop is given by
\begin{align}
    &
    \begin{tikzpicture}[scale=0.5, baseline={(current bounding box.center)}]
      \draw[decorate, decoration={snake, segment length=4pt, amplitude=2pt}] (0,0) circle (1cm);
      \draw[cross] (0,1) circle (0.3cm);
     \node[below] at (0,-1.2) {};
    \end{tikzpicture}
    =
    \nonumber \\
    & =
    \frac{1}{2}\mathcal{V}_2 \sum_{n=-\infty}^{\infty}
    \int \frac{\mathrm{d}^2 p}{(2\pi)^2}
    \vec{p}^{\,2}
    \frac{\left[\partial_t r_b (t, p)\right]}{\omega_n^2 + \Vec{p}^{\,2} \left[1 + r_b(t, p)\right] + \partial_\sigma^2 U(t, \sigma)}
    \nonumber \\
    & = -
    \mathcal{V}_2
    \beta
    \frac{k^4}{
    8 \pi E_b (t, \sigma)
    }
    \left[
        1 + 2 n_b \left(\beta E_b (t, \sigma)\right)
    \right]
    \;.
\end{align}
For the fermion loop we get
\begin{align}
    & -2\;
    \begin{tikzpicture}[scale=0.5, baseline={(current bounding box.center)}]
      \draw (0,0) circle (1cm);
      \draw[->] (0,-1) -- (-0.01,-1);
      \draw[fill=white, cross] (0,1) circle (0.3cm);
     \node[below] at (0, -1.2) {};
    \end{tikzpicture}
    =
    \nonumber \\
    &
    =
    -
    \onehalf
    d_\gamma
    N_f
    \mathcal{V}_2 \nonumber \\
    & \times 
    \sum_{n=-\infty}^{\infty}
    \int \frac{\mathrm{d}^2 p}{(2\pi)^2}
    \frac{
        p^2
        \left[\partial_t r_b(t, p)\right]
    }{
            \left(\nu_n + i \mu\right)^2 
            +
            \Vec{p}^{\,2} \left[1 + r_b (t, p)\right]
            +
            m_f^2
    }
    \nonumber
    \\
    &
    =
    d_\gamma
    N_f
    \mathcal{V}_2
    \beta
    \frac{
        1
    }{
        8\pi
    }
    \frac{
        k^{4}
    }{
         E_f(t, \sigma)
    }
    \nonumber \\
    & 
    \times
    \left[
        1 
        -
        n_f \left(\beta \left[E_f(t, \sigma) + \mu\right]\right)
        -
        n_f \left(\beta \left[E_f(t, \sigma) - \mu\right]\right)
    \right]
    \;.
\end{align}

After rescaling the scale-dependent effective potential and the scalar field as
\begin{align}
    \sigma \rightarrow& \sqrt{N_f}\, \sigma \;, \\
    U(t, \sigma) \rightarrow& N_f \, U(t, \sigma) \;,
\end{align}
the flow equation for the rescaled effective potential reads
\begin{align}
    &
    \partial_t U(t, \sigma)
    \nonumber\\
    =&
    -
    \frac{d_\gamma }{8 \pi}
    \frac{k^{4}}{ N_f E_b (t, \sigma)}
    \left[
        1 + 2 n_b \left(\beta E_b(t, \sigma)\right)
    \right]
    \nonumber\\
    &+\frac{d_\gamma }{8 \pi}
    \frac{k^{4}}{E_f (t, \sigma)} \nonumber \\
& \times     \left[
        1
        -
        n_f \left(\beta\left[E_f (t, \sigma)+ \mu\right] \right)
        -
        n_f \left(\beta\left[E_f (t, \sigma)- \mu\right]\right)
    \right]
    \;.
\end{align}

\subsection{Nonzero Magnetic Field}

In this model, only fermions carry electric charge. 
Therefore, for nonvanishing magnetic field, only the calculation of the fermion loop changes. 
To introduce the magnetic field we employ the replacements from Eqs.~(\ref{eq:momentumReplacement}), (\ref{eq:volumeIntegralToLandauSum}),
    \begin{align}
        &
        -2 \;
        \begin{tikzpicture}[scale=0.5, baseline={(current bounding box.center)}]
          \draw (0,0) circle (1cm);
          \draw[->] (0,-1) -- (-0.01,-1);
          \draw[fill=white, cross] (0,1) circle (0.3cm);
         \node[below] at (0, -1.2) {};
        \end{tikzpicture}
        \nonumber\\
        &
        =
        d_\gamma N_f \mathcal{V}_2 
        \frac{|q B|}{4\pi}
        \sum_{n=-\infty}^{\infty}
        \sum_{l=0}^{\infty}
        \alpha_l
        \frac{
            k^2
            \theta\left(k^2 - 2|qB|l\right)
        }{
            \left(\nu_n + i \mu\right)^2
            +
            k^2 
            +
            m_f^2 (t, \sigma)
        }
        \nonumber
        \\
        &
        =
        d_\gamma N_f \beta \mathcal{V}_2 
        \frac{|qB|}{8\pi}
        \frac{
            k^2       
        }{E_f(t, \sigma)}\, \left[
                1
                +
                2 N_{LL}(k^2)
            \right]
        \nonumber \\
        & \times
        \left[
            1
            -
            n_f \left(\beta \left[E_f(t, \sigma) + \mu \right]\right)
            -
            n_f \left(\beta \left[E_f(t, \sigma) - \mu \right]\right)
        \right]
        \;,
\end{align}
where the Heaviside function gives an upper bound on the sum over Landau levels. 
Since this sum is independent of the summation index, the calculation can be trivially performed. 
The final step is then the evaluation of the Matsubara sum.

\section{Numerics}
\label{sec:numerics}
In this appendix, we discuss the numerical method used to solve the flow equations derived in Appendix~\ref{Flow-Equations}. 
\subsection{FRG Flow Equation in Conservative Form}

Recently it has been found that the flow equation for the effective potential can be cast into conservative form~\cite{Grossi:2019urj}. 
This allows to use methods from numerical hydrodynamics to solve the flow equations.
Here, we employ the Kurganov-Tadmor (KT) scheme \cite{KURGANOV2000241}, a high-resolution finite-volume method for solving hyperbolic partial differential equations, which solves the Riemann problem approximately at each cell boundary.
In this way, we obtain a numerically stable solution even in the presence of shock waves created by a nonzero chemical potential during the FRG flow~\cite{Steil:2021cbu,Stoll2021arXiv210810616S}.
This allows us to go deeper into the IR regime and, consequently, to resolve the phase diagram to higher accuracy than with other methods. 
For a more detailed discussion of the numerical implementation, we refer to Ref.~\cite{Stoll2021arXiv210810616S}.

Following Ref.~\cite{Stoll2021arXiv210810616S} we take the partial derivative of the flow equation with respect to $\sigma$ and define $u (t, \sigma)= \partial_\sigma U_t (t, \sigma)$. 
The resulting flow equation has the form of a conservation law,
\begin{align} \label{eq:cons}
    \partial_t u (t, \sigma)
    & =
    \partial_\sigma Q(t, \partial_\sigma u)
    +
    S(t, \sigma)
    \;,
\end{align}
where $u(t, \sigma)$ is the conserved quantity and $Q(t, \partial_\sigma u)$ is a non-linear diffusion flux, while $S(t, \sigma)$ is a source/sink term.

Since the GNY model does not contain Goldstone modes, the conservation equation~\eqref{eq:cons} does not feature an advection term.
Without such a term, the KT scheme reduces to
\begin{align}
    \frac{d}{dt} u_j (t)
    =
    &
    \frac{
        P_{j+1/2}(t)
        -
        P_{j-1/2}(t)
    }{\Delta \sigma}
    + 
    S(t, \sigma_j)
    \;,
\end{align}
where $P_{j+1/2}(t)$ are the discretized diffusion fluxes given by
\begin{align}
    P_{j+1/2}
    =
    &
    \frac{1}{2}
    \left[
        Q\left(u_j , \frac{u_{j+1} - u_j }{\Delta \sigma} \right)
        +
        Q\left(u_{j+1} , \frac{u_{j+1} - u_j }{\Delta \sigma} \right)
    \right]
    \;.
\end{align}

In our case the diffusion flux does not depend on $u(t, \sigma)$ itself, but only on its derivatives, leading to further simplifications.
The KT scheme converts the flow equation, which is a partial differential equation, into a set of ordinary coupled differential equations. 
We can then solve these equations using standard methods to integrate them over RG time.
Specifically we are using \texttt{solve\_ivp} with \texttt{LSODA} as the method from the \textit{SciPy} library \cite{2020SciPy-NMeth}. 
We use relative and absolute tolerances of $10^{-8}$ for the adaptive time-stepping. 
The flow is integrated from $t=0$ (UV cutoff $\Lambda=10^3$) down to $t_{\mathrm{IR}} = \ln(\Lambda/k_{\mathrm{IR}})$ with $k_{\mathrm{IR}}=10^{-2}$.

\subsection{Initial and Boundary Conditions}

The initial condition (UV limit) for the effective potential is given by its classical value in vacuum,
\begin{equation}
    \label{eq:mfUVPotential}
    U_\Lambda (\sigma)
    =
    \frac{(h\sigma)^2}{2 g^2}
    \;.
\end{equation}
We set $h=1$, and measure all dimensionful quantities in units of $\sigma_0$.
We fit the parameter $g^2$ using a Newton-Raphson root-finding algorithm so that the resulting effective potential in the IR has a minimum at $\sigma_0 = 1$.

The $\sigma$ field is discretized by equally spaced finite-volume cells of size $\Delta \sigma$ in the domain $[0, \sigma_{\text{max}}]$. For the implementation
of the KT scheme, we also need ghost cells. These are placed outside the boundaries at $\sigma_{-1} = -\Delta \sigma$ and $\sigma_{n+1} = \sigma_{\text{max}} + \Delta \sigma$ 
respectively. We use the following parameters for our numerical implementation:
\begin{center}
    \begin{tabular}{|c|c|c|c|}
        \hline 
        $\Lambda$ & $k_{\mathrm{IR}}$ & $\sigma_{\mathrm{max}}$ & $\Delta \sigma$ \\
        \hline
        $10^3$ & $10^{-2}$ & $5$ & $5 \cdot 10^{-3}$ \\
        \hline
    \end{tabular}
\end{center}

\bibliography{refs}

\end{document}